\documentclass[aps,prr,showpacs,twocolumn,superscriptaddress]{revtex4-1}
\usepackage{graphicx}
\usepackage{epstopdf}
\usepackage{dcolumn}
\usepackage{bm}
\usepackage[utf8x]{inputenc}
\usepackage{mathrsfs}
\usepackage{amssymb}
\usepackage{amsmath}
\usepackage{color}
\usepackage{xcolor}
\usepackage{natbib}
\begin{document}
	\title{Applications of deep learning to relativistic hydrodynamics}
	
\author{Hengfeng Huang}
\affiliation{Department of Physics and State Key Laboratory of Nuclear Physics and Technology, Peking University, Beijing 100871, China}
\affiliation{Collaborative Innovation Center of Quantum Matter, Beijing 100871, China}

\author{Bowen Xiao}
\affiliation{Institute of Computer Science and Technology, Peking University, Beijing 100080, China}

\author{Ziming Liu}
\affiliation{Department of Physics and State Key Laboratory of Nuclear Physics and Technology, Peking University, Beijing 100871, China}

\author{Zeming Wu}
\affiliation{Department of Physics and State Key Laboratory of Nuclear Physics and Technology, Peking University, Beijing 100871, China}
\affiliation{Collaborative Innovation Center of Quantum Matter, Beijing 100871, China}

\author{Yadong Mu}
\affiliation{Institute of Computer Science and Technology, Peking University, Beijing 100080, China}
\affiliation{{Center for Data Science}, Peking University, Beijing 100871, China}

\author{Huichao Song}
\affiliation{Department of Physics and State Key Laboratory of Nuclear Physics and Technology, Peking University, Beijing 100871, China}
\affiliation{Collaborative Innovation Center of Quantum Matter, Beijing 100871, China}
\affiliation{Center for High Energy Physics, Peking University, Beijing 100871, China}

	\date{\today}
	\begin{abstract}
Relativistic hydrodynamics is a powerful tool to simulate the evolution of the quark gluon plasma (QGP) in relativistic heavy ion collisions.  Using 10000 initial and final profiles generated from 2+1-d relativistic hydrodynamics {\tt VISH2+1}  with {\tt MC-Glauber} initial conditions, we train a deep neural network based on {\tt stacked U-net}, and use it to predict the final profiles associated with various initial conditions, including {\tt MC-Glauber}, {\tt MC-KLN}, {\tt AMPT} and {\tt TRENTo}. A comparison with the {\tt VISH2+1} results shows that the network  predictions can nicely capture the magnitude and inhomogeneous structures of the final profiles, and nicely describe the related eccentricity distributions $P(\varepsilon_n)$ (n=2, 3, 4). These results indicate that deep learning technique can capture the main features of the non-linear evolution of hydrodynamics, showing its potential to largely accelerate the event-by-event simulations of relativistic hydrodynamics.
	\end{abstract}
	\maketitle

\section{Introduction}

\begin{figure*}[htbp]
	\centerline{\includegraphics[width=0.9\linewidth,height=1.6in]{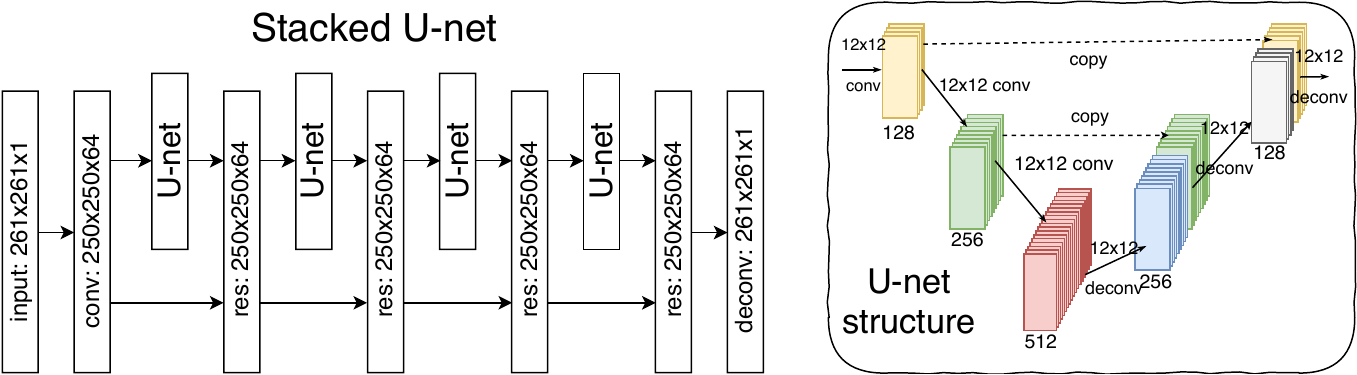}}
	\caption{ An illustration of the encode-decode network, {\tt stacked U-net}, which consists of the input and out layers and four residual U-net blocks. The right figure shows the U-net structure, and the depth of the hidden layer is written on the top of them. }
	\label{net}
\end{figure*}

In recent years deep learning~\cite{Goodfellow-et-al-2016,lecun2015deep,silver2016mastering} has achieved great success in both daily life and in sciences. In particular, deep learning methods have been implemented to various research areas in physics, including the search of gravitational lens\cite{hezaveh2017fast,Petrillo:2017njm}, identifying and classifying the phases of Ising model~\cite{sun2018deep,carrasquilla2017machine,van2017learning,broecker2017machine}, solving the quantum many body problem~\cite{carleo2017solving,gao2017efficient}, etc. In high energy physics, it has been applied to the search of Higgs and exotic particles~\cite{baldi2014searching,baldi2016parameterized}, classification of jet structure~\cite{Cogan:2014oua,baldi2016jet,chakraborty2019interpretable}, etc. In the field of relativistic heavy-ion collisions,machine learning and deep neural networks have been employed to attack the problems of identifying the equation of state of hot QCD matter~\cite{pang2018equation}, jet-flavor classification in heavy-ion collisions~\cite{chien2019probing}, distinguishment between spinodal and Maxwell first-order phase transition~\cite{steinheimer2019machine}, detecting nuclear shape deformation~\cite{pang2019interpretable}, Bayesian extraction of transport properties of the hot QCD matter~\cite{Bernhard:2016tnd,Bernhard:2019bmu,Everett:2020yty,Everett:2020xug}, the phase diagram of two-dimensional complex scalar field theory~\cite{zhou2019regressive}, and principal component analysis of collective flow~\cite{PhysRevLett.114.152301, mazeliauskas2015subleading, sirunyan2017principal, liu2019principal, Liu:2020ely,PhysRevC.100.054905, Altsybeev_2020}.

In this paper, we will apply deep learning to relativistic hydrodynamics, which is a useful tool to simulate the  macroscopic evolution of relativistic systems in high energy nuclear physics and astrophysics~\cite{landau2013course}. Relativistic hydrodynamics solves the transport equations of the energy momentum tensor and charge currents based on the conservation laws. In relativistic heavy ion collisions, it has nicely described and predicted various flow data of the quark-gluon plasma (QGP), which played an important role in the discovery of the strongly coupled QGP and its nearly perfect fluid nature~\cite{gyulassy2005new,muller2006results,huovinen2004hydrodynamical,kolb2004hydrodynamic,
heinz2013collective,gale2013hydrodynamic,song2017collective,muller3233first,Song:2010mg,Schenke:2010rr,Gale:2012rq,
Bernhard:2016tnd,Zhu:2016puf,Zhao:2017yhj,Bernhard:2019bmu,Everett:2020yty,Everett:2020xug}. However, traditional hydrodynamic simulations are time consuming. For example, the calculation of various flow harmonics requires $\sim$ 1000 event-by-event hydrodynamic simulations, which takes $\sim$500 and $\sim$10000 cpu hours for the typical 2+1-d and 3+1-d simulations, respectively~\cite{heinz2013collective,gale2013hydrodynamic,song2017collective,shen2016iebe}.
Basically, relativistic hydrodynamics translate the initial conditions to final profiles through solving a set of non-linear differential equations. In this work, we will explore whether the deep neural network could capture the main features of the non-linear evolution of 2+1-d hydrodynamics, and the possibilities to accelerate the related event-by-event simulations.  {Close to this work are Bayesian emulators~\cite{Pratt:2015zsa,Everett:2020xug}, which are powerful in constraining equation of states and transport coefficients, yet are not designed to predict the whole profiles of energy density and flow velocity.}

It is worthwhile to mention that the interdisciplinary contributions of this work are two-fold: from the physics perspective, we  speed up hydrodynamic simulations times with a deep neural network, while still capturing the details for the final profiles of the expanding QGP; On the other hand from the machine learning angle, we highlight the expressive power of the {\tt stacked U-net} model, as well as its ability to approximate partial differential equation (PDE) in this particular task of relativistic hydrodynamics.

The paper is organized as follows: In Section II, we introduce the relativistic hydrodynamics and network design. In Section III, we show the results obtained from the network, followed by the discussions and conclusions in Section IV.

\section{Models}

\subsection{Relativistic hydrodynamics}
\par In this paper,  we focus on relativistic ideal hydrodynamics with zero viscosity and charge densities, which solves the transport equations of the energy momentum tensor $T^{\mu\nu}$:
	\begin{equation}
		\partial_{\mu}T^{\mu\nu}=0
	\end{equation}
where $T^{\mu \nu}=(e+p) u^{\mu}u^{\nu}-p g^{\mu\nu}$,
$e$ is the energy density, $p$ is the pressure, and $u^\mu$ is the four velocity with $u^\mu u_\mu=1$. With an assumption
of longitudinal boost invariance, we solve the 2+1-dimensional hydrodynamic equations with an ideal EoS $p=\frac{e}{3}$, using the code {\tt VISH2+1}~\cite{shen2016iebe,song2008causal}~\footnote{In $(\tau,x,y,\eta)$ coordinate ($\tau=\sqrt{t^2-z^2}$ and $\eta=\frac{1}{2}\mathrm{ln}\frac{t+z}{t-z}$), the energy density and pressure from 2+1-d hydrodynamics are longitudinal boost-invariant without a dependence on $\eta$, $e$=$e(\tau,x,y)$ and $p$=$p(\tau,x,y)$. Correspondingly, the four flow velocity are expressed as $\gamma_\bot(1, v^x(\tau,x, y), v^y(\tau,x, y), 0)$  with $v^\eta=0$~\cite{huovinen2004hydrodynamical,kolb2004hydrodynamic}.}. The initial energy density profiles can be generated by some initial condition models, such as {\tt MC-Glauber}~\cite{miller2007glauber,hirano2009eccentricity}, {\tt MC-KLN}~\cite{drescher2007effects,drescher2007eccentricity,hirano2009eccentricity}, {\tt AMPT}~\cite{pang2012effects,xu2016high,Zhao:2017rgg} and {\tt TRENTo}~\cite{moreland2015alternative} with zero initial transverse flow velocity.   We run {\tt VISH2+1} with three selected fixed evolution times $\tau-\tau_0=2.0$, 4.0 and 6.0 fm/c ($\tau_0$=0.6 fm/c) to obtain the  energy momentum tensor $T^{\tau\tau}(\tau, x, y)$, $T^{\tau x}(\tau, x, y)$, $T^{\tau y}(\tau, x, y)$ profiles at these times.  For numerical accuracy, the time step and grid sizes of the simulations are set to $d \tau=0.04 \ \mathrm{fm/c}$ and $d x = d y =0.1 \ \mathrm{fm}$, within a fixed transverse area of $13 \ \mathrm{fm} \times 13 \ \mathrm{fm}$ that have been used to describe the typical QGP expansion in relativistic heavy ion collisions.

\begin{figure*}[htbp]
	\centerline{\includegraphics[width=0.92\linewidth, height=5.0in]{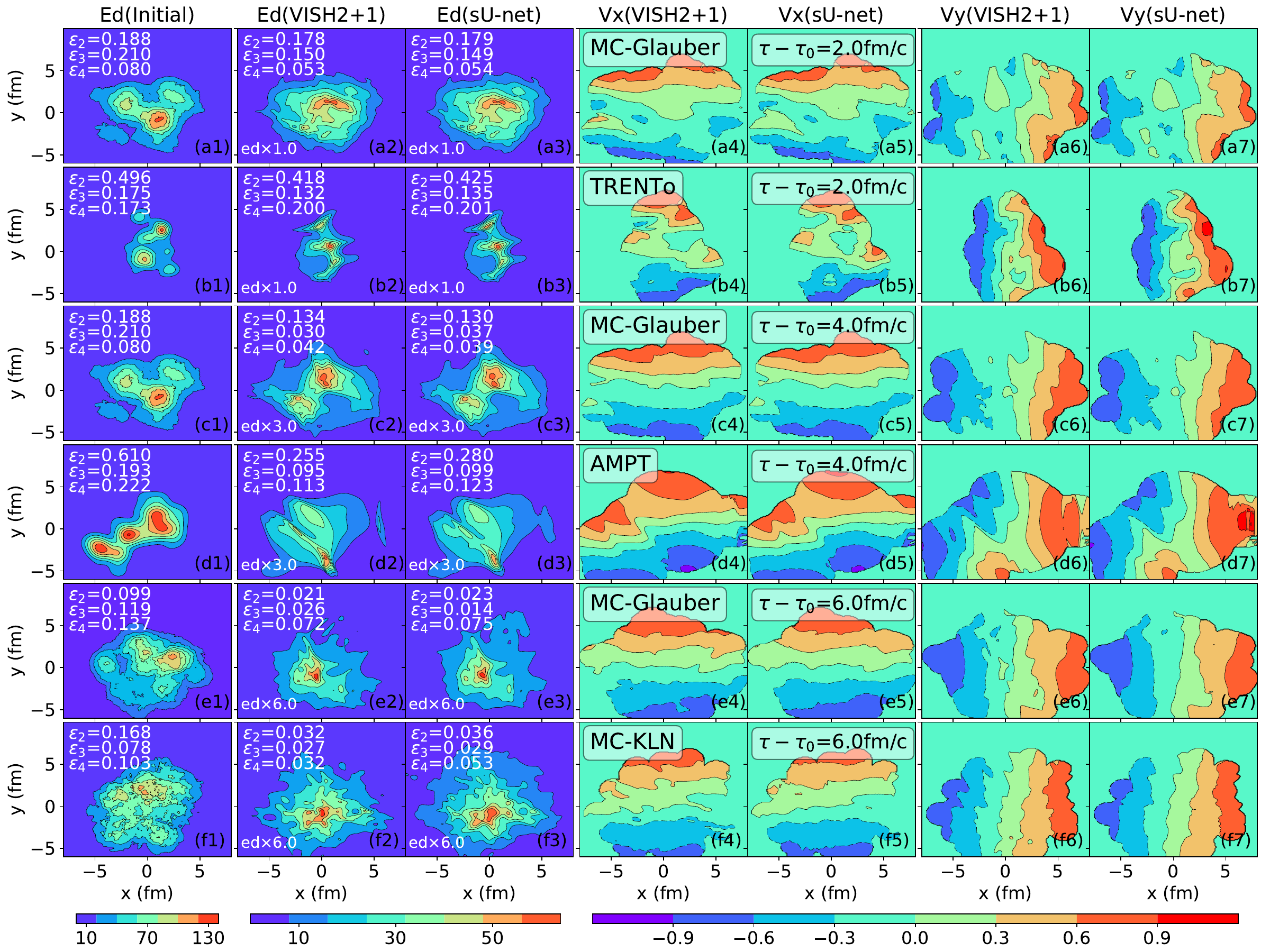}}
	\caption{Energy density and flow velocity profiles at $\tau-\tau_0=2.0, \ 4.0,  \ 6.0   \ \mathrm{fm/c}$, calculated from {\tt VISH2+1} and predicted by the network for six test cases with initial profiles generated from {\tt MC-Glauber},  {\tt MC-KLN},
		{\tt AMPT} and {\tt TRENTo}.}\label{cmp}
\end{figure*}

\subsection{Network design}
For deep learning,  the initial and final energy momentum tensor $T^{\tau\tau}$, $T^{\tau x}$, $T^{\tau y}$ profiles from hydrodynamics are treated as initial and final image sets with $261\times261$ pixels.  In practice, we first run the event-by-event hydrodynamic simulations to obtain 10000 initial and final image sets, then use them to train the deep neural network, which aims at achieving nice predictions of the final energy density and flow velocity profiles for other input initial conditions.

The related network we adopted in this work is the {\tt stacked U-net} ({\tt sU-net})~\cite{he2016deep}, which is a variation of the traditional encoder-decoder network that could enhance gradient flow in the deeper part of the network during back propagation. Fig.~1 presents an intuitional view of the network structure. It consists of 4 serially connected U-net blocks with residual connections between them. Each U-net block has 3 convolution layers and 3 deconvolution layers. In each U-net block, the output of the first two convolution layers are also fed into the last two deconvolution layers respectively by concatenating the feature maps along the channel dimension. The activation function for all layers except for the output one is \emph{Leaky ReLU} $f\left( x \right) = \max \left\{ {x,0.03x} \right\}$, while that for the output layer is \emph{softplus} $f\left( x \right) = \ln \left( {1 + {e^x}} \right)$ for $T^{\tau\tau}$ mapping and $f\left( x \right)=x$ for $T^{\tau x}$ and $T^{\tau y}$ mapping. To make the network focus more on local patterns, we set the kernel size of all convolution and deconvolution layers to $3\times3$. The loss function of the network is \emph{normalized MAE loss} $Loss = \frac{ |y_1 - y_0|}{ | y_0 |}$, where $y_1$ is the output of the network and $y_0$ is the ground truth.
We use the standard mini-batch stochastic gradient descent algorithm for optimization. The batch size for training is 16 and learning rate exponentially decays from $10^{-3}$ to $10^{-5}$. {Each weight is randomly initialized from the uniform distribution on $[-0.001,0.001]$ and each bias is set to zero.} Our code is built with TensorFlow and the training process runs for about 1 day on a machine with single NVIDIA Tesla P40 GPU, using 10000 ``initial" and ``final" profiles from {\tt VISH2+1} hydrodynamic simulations.

We have noticed that, although one trained {\tt sU-net} can make nice predictions for shorter hydrodynamic evolution, it fails to accurately predict the final profiles of longer evolution time ($\tau-\tau_0>4.0 \ \mathrm{fm/c}$) from the initial profiles at $\tau_0$. Considering that the evolving QGP system is highly non-linear and tends to smear-out its initial structures during longer evolution,
we divide the whole evolution time $\tau-\tau_0$ into $n$ parts with equal time interval $\Delta \tau$: $\tau-\tau_{n-1}$ ... $\tau_2-\tau_1$, $\tau_1-\tau_0$. For each evolution part, we train an individual {\tt sU-net} using the corresponding ``initial" and ``final" profiles from hydrodynamics.  To predict the final profiles at $\tau$ from initial profiles at $\tau_0$, we first use the trained {\tt sU-net-1} to predict the profiles at time $\tau_1$ and then use them as the initial conditions for {\tt sU-net-2} to predict the profiles at time $\tau_2$  and so on. In this way, the combined {\tt sU-net} series (i=1...n) mimic the hydrodynamic evolution with much larger time step $\Delta \tau$ that can not be managed by traditional hydrodynamic  algorithm (In more detail, for the following evolution with $\tau-\tau_0= 6.0 \ \mathrm{fm/c}$, we set n=3 with $\Delta \tau = 2.0 \ \mathrm{fm/c}$). Note that {\tt sU-net-1}, {\tt sU-net-2} and {\tt sU-net-3} are not identical since the initial and final profiles are generated by {\tt VISH2+1}, which implement the 2+1-d hydrodynamic equations explicitly depended on $\tau$~\cite{Song:2009gc,song2008causal}.
% It also helps to significantlyaccelerate the related event-by-event hydrodynamic simulations. %Please refer to the following long paper for more details~\cite{Huang}.

\section{Results}

\begin{figure*}[htbp]
	\centerline{\includegraphics[width=1\linewidth]{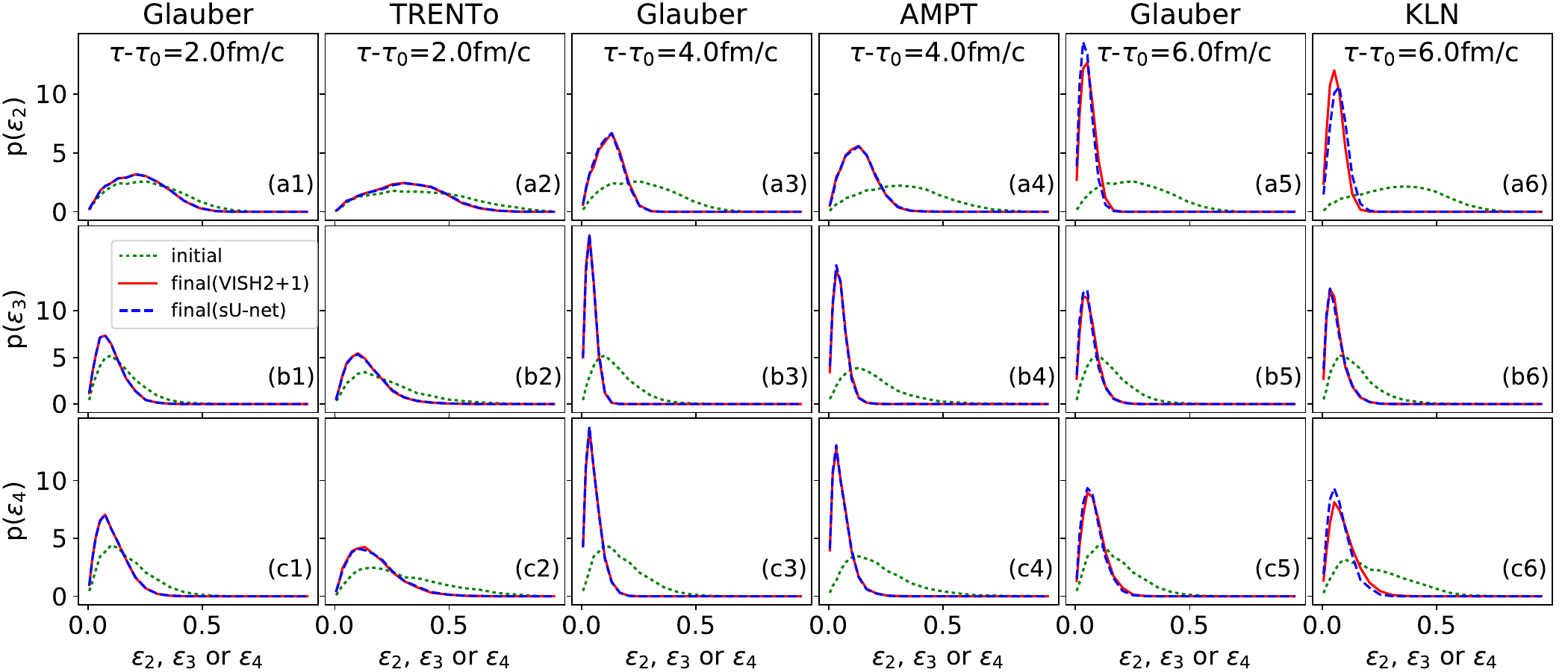}}
	\caption{ Eccentricity distribution $P(\varepsilon_n)$ (n=2, 3, 4), at $\tau-\tau_0=2.0, \ 4.0,  \ 6.0  \ \mathrm{fm/c}$, calculated from {\tt VISH2+1} and predicted by the network for 100000 tested initial profiles generated from {\tt MC-Glauber}, {\tt MC-KLN}, {\tt AMPT} and {\tt TRENTo}.}\label{eccentricity}
\end{figure*}

As explained in the above text, we first use 10000 initial and final image sets from {\tt VISH2+1} with {\tt MC-Glauber} initial conditions to train the combined {\tt stacked U-net}, and then use the trained network to predict the final profiles from the initial profiles generated from {\tt MC-Glauber}, {\tt MC-KLN}, {\tt AMPT} and {\tt TRento} as tests.  Fig.~2 presents a comparison between the results from {\tt VISH2+1} hydrodynamic evolution and the predictions from the network at $\tau-\tau_0=2.0, \ 4.0,  \ 6.0   \ \mathrm{fm/c}$ for 6 selected test cases. It shows that the well designed and trained network could nicely predict the
final states, which captures the structures of the contour plots for both final energy density and flow velocity. It is impressive that, although the network is trained with the initial and final image sets associated with  the {\tt MC-Glauber} initial conditions, it could still nicely predict the final profiles of other initial conditions with different fluctuation patterns, as shown in panel (b), (d) and (f).

To further evaluate the predictive power of the network, we further calculate the eccentricity coefficients
\begin{equation}
\varepsilon_n= \frac{\int r dr d\phi \  r^{n} e(r,\phi) e^{in\phi}}{ \int  r dr d\phi \  r^{n} e(r,\phi)} \ \ \ (n=2, 3, 4)
\end{equation}
for initial and final energy density $e(r,\phi)$ profiles, which are quantities commonly used
to evaluate the deformation and inhomogeneity of the QGP fireball in relativistic heavy ion collisions~\cite{heinz2013collective,gale2013hydrodynamic,song2017collective,shen2016iebe}.
These values of $\varepsilon_n$ (n=2, 3, 4) for these 6 selected test cases are written in the related panels~(a-f). From Fig.~2 and the calculated values of $\varepsilon_n$ (n=2, 3, 4),  we have also noticed that differences between the hydrodynamic results and the network predictions increase for longer evolution
time since the combined {\tt sU-net} series tend to accumulates errors with more sU-net added. %For details, please refer to~\cite{Huang}.

Fig.~3 presents the eccentricity distributions $P(\varepsilon_n)$
for the energy density profiles at evolutions times $\tau-\tau_0 =$ 2.0 fm/c, 4.0 fm/c and
6.0 fm/c, calculated from {\tt VISH2+1} and predicted
from the network for 10000 tested initial profiles generated from {\tt MC-Glauber}, {\tt MC-KLN}, {\tt AMPT}
and {\tt TRENTo}. For all these tested cases, the final eccentricity distributions $P(\varepsilon_n)$
(n=2, 3, 4) from the network almost overlap with the ones from {\tt VISH2+1}, which also
obviously deviate from the initial eccentricity distributions $P_0(\varepsilon_n)$. In Fig.~4, scatter plots show event-by-event comparisons between true eccentricities of `final' profiles and predicted ones, and histograms of errors are plotted in the inset figure.

We also find that, with the well trained network, the final state profiles can be speedily generated from the initial profiles. Compared with the 10-20 minute calculation time with traditional CPU for a single-event hydrodynamic evolution, the network takes several seconds to directly generate the final profile for different types of initial profiles with the P40 GPU,
which shows the potential to accelerate the realistic event-by-event hydrodynamic simulations in the near future. {However, given the fact that 50-100x speedup of hydro simulations can be already achieved by switching from CPU to GPU~\cite{Pang:2018zzo,Bazow:2016yra}, we believe there is still much room to improve our proof-of-concept first step in further studies.}

\begin{figure*}[htbp]
	\centerline{\includegraphics[width=0.95\linewidth]{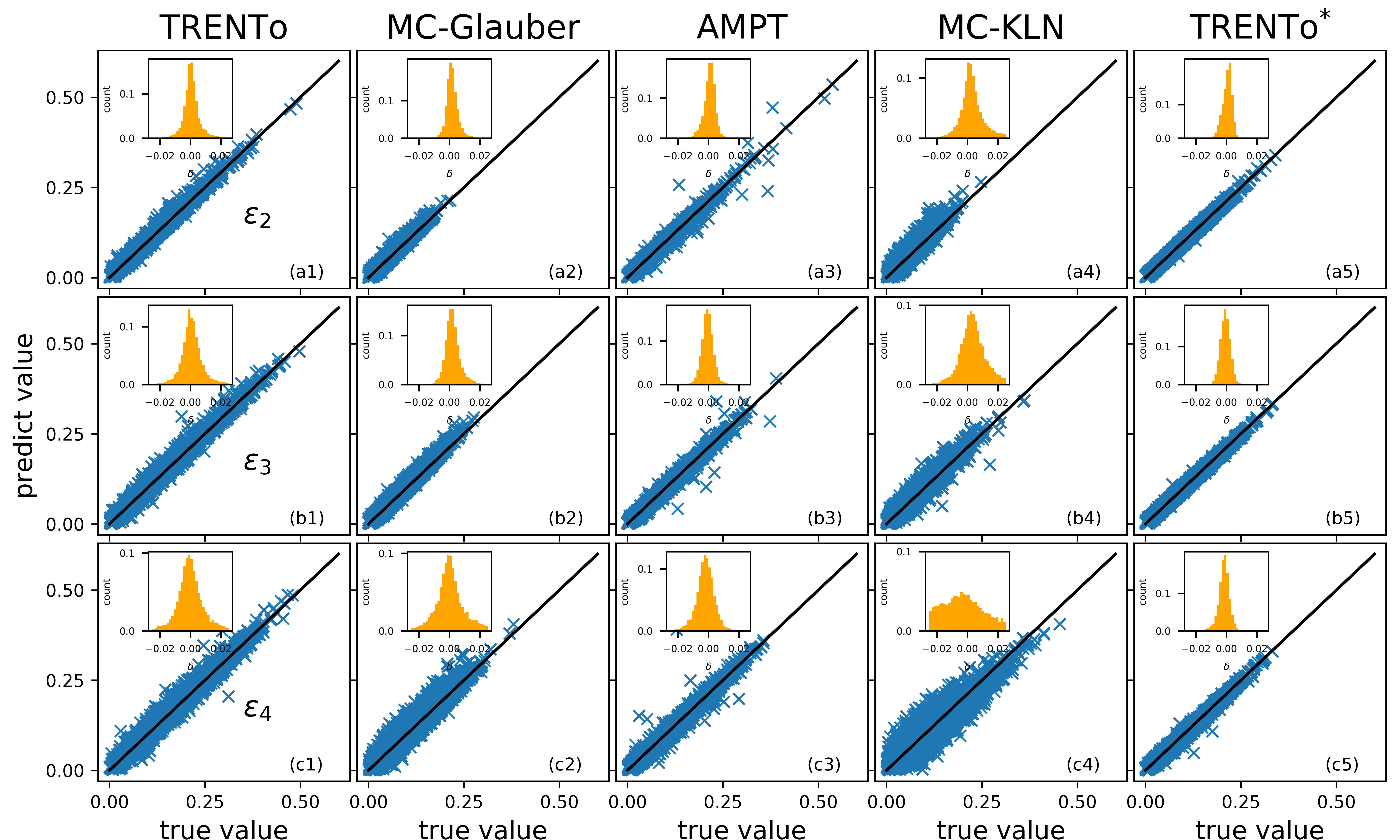}}
	\caption{ Event-by-event camparisons of eccentricities $\varepsilon_n$ (n=2, 3, 4) at $\tau-\tau_0=6.0\ \mathrm{fm/c}$, calculated from {\tt VISH2+1} and predicted by the network for 100000 tested initial profiles generated from {\tt MC-Glauber}, {\tt AMPT}, {\tt MC-KLN}, and {\tt TRENTo} (with two set of parameters distinguished by *).}\label{ebe_predict}
\end{figure*}

\section{Discussion and Conclusion}

Using 10000 initial and final energy momentum tensor profiles from {\tt VISH2+1} hydrodynamics with {\tt MC-Glauber} initial conditions, we successfully trained a deep neural network based on {\tt stacked U-net}, and use it to predict the final profiles for different initial conditions, including {\tt MC-Glauber}, {\tt MC-KLN}, {\tt AMPT} and {\tt TRENTo}.  A comparison with the {\tt VISH2+1} results showed that the network predictions could
nicely capture the magnitude and inhomogeneous structures of the final profiles, which also
nicely describe the related eccentricity distributions $P(\varepsilon_n)$ (n=2, 3, 4). These results indicate that deep learning could capture the main feature of the non-linear evolution of hydrodynamics, which also shows the potential of largely accelerating the realistic event-by-event hydrodynamic simulations in relativistic heavy ion collisions.

In order to outline the highlights, as well as point out limitations of this work, we further explain the following characteristics that mark good works and provide guidelines for future studies.
	
	\par {\bf Universality}: Deep learning might not learn the realistic physics underlying the dataset. By contrast, sometimes its predictions are based on non-physical features in the dataset, as has been pointed out in~\cite{zhang2018machine}. In this work, we exclude such undesirable possibility by training our deep model on {\tt MC-Glauber} initial conditions and test on results for other initial models including {\tt MC-KLN}, {\tt AMPT} and {\tt TRENTo}.
\par {\bf Causuality}: Due to speed of light as an upper bound for all physical speeds, our neural network should satisfy such causual relations, otherwise it will produce non-physical results. The joint use of convolutional layers and the stacked structure elegantly handles this issue by allowing one pixel to influence its neighborhoods only. Such causuality in convolutional layers is known as receptive field~\cite{luo2016understanding} in the machine learning literature. {More concretely suppose our CNN has $L$ layers with the $i$-th layer using convolutional filters of size $(2n_i+1)\times (2n_i+1)$, then it is reasonable to match the size of the receptive field $R_r=(\sum_i n_i)\Delta x$ ($\Delta x$ is the grid length) with the size of light cone $ R_l=c(\tau-\tau_0)$. If $R_r<R_l$, the expressivity of the neural network is bottlenecked; If $R_r<R_l$, the neural network is unnecessarily expressive which might lead to longer training time.}
	\par {\bf Utility}: One limitation of this work is the fixed time output. Future works will consider more flexible architectures (e.g. physics-informed neural network in~\cite{raissi2019physics}) to obtain the energy-momentum tensor at the freeze-out surface with more realistic implementation in heavy ion collisions.
	\par {\bf Interpretability}: Another minor limitation of the {\tt stacked U-net} model lies in the lack of interpretability. In future works, we will investigate the possibilities of encoding physics explicitly in network design, as in~\cite{raissi2019physics}. Efforts on gaining interpretability of deep learning in heavy-ion collisions include~\cite{pang2019interpretable,pang2018equation}.

In summary, our current investigations mainly focus on mimicking the 2+1-dimensional hydrodynamic evolution with fixed evolution time,  using the deep learning technique. On the one hand, for more realistic implementation to relativistic heavy ion collisions, it is worthwhile to explore the possibilities of mapping the initial profiles to the final profiles on the freeze-out surface with fixed energy density as well as extending the related investigations to 3+1-dimensional simulations. On the other hand, it is also worthwile to develop computational tools that are more transparent for scientific evulations, where a possible way is to encode physical prior (the functional form of the PDE, boundary contions etc.) into the network architecture. \\[0.20in]

\section{Acknowledgments}
We thanks the discussions from  L.~G.~Pang, K.~Zhou and X.-N Wang.  H.~H, Z.~L and H.~S. are supported by the NSFC under grant No. 11675004 and No. 12075007. B.~X. and Y.~M. are supported by NSFC under grant no. 61772037.

\bibliography{ref}

\begin{thebibliography}{65}%
\makeatletter
\providecommand \@ifxundefined [1]{%
 \@ifx{#1\undefined}
}%
\providecommand \@ifnum [1]{%
 \ifnum #1\expandafter \@firstoftwo
 \else \expandafter \@secondoftwo
 \fi
}%
\providecommand \@ifx [1]{%
 \ifx #1\expandafter \@firstoftwo
 \else \expandafter \@secondoftwo
 \fi
}%
\providecommand \natexlab [1]{#1}%
\providecommand \enquote  [1]{``#1''}%
\providecommand \bibnamefont  [1]{#1}%
\providecommand \bibfnamefont [1]{#1}%
\providecommand \citenamefont [1]{#1}%
\providecommand \href@noop [0]{\@secondoftwo}%
\providecommand \href [0]{\begingroup \@sanitize@url \@href}%
\providecommand \@href[1]{\@@startlink{#1}\@@href}%
\providecommand \@@href[1]{\endgroup#1\@@endlink}%
\providecommand \@sanitize@url [0]{\catcode `\\12\catcode `\$12\catcode
  `\&12\catcode `\#12\catcode `\^12\catcode `\_12\catcode `\%12\relax}%
\providecommand \@@startlink[1]{}%
\providecommand \@@endlink[0]{}%
\providecommand \url  [0]{\begingroup\@sanitize@url \@url }%
\providecommand \@url [1]{\endgroup\@href {#1}{\urlprefix }}%
\providecommand \urlprefix  [0]{URL }%
\providecommand \Eprint [0]{\href }%
\providecommand \doibase [0]{http://dx.doi.org/}%
\providecommand \selectlanguage [0]{\@gobble}%
\providecommand \bibinfo  [0]{\@secondoftwo}%
\providecommand \bibfield  [0]{\@secondoftwo}%
\providecommand \translation [1]{[#1]}%
\providecommand \BibitemOpen [0]{}%
\providecommand \bibitemStop [0]{}%
\providecommand \bibitemNoStop [0]{.\EOS\space}%
\providecommand \EOS [0]{\spacefactor3000\relax}%
\providecommand \BibitemShut  [1]{\csname bibitem#1\endcsname}%
\let\auto@bib@innerbib\@empty
%</preamble>
\bibitem [{\citenamefont {Goodfellow}\ \emph {et~al.}(2016)\citenamefont
  {Goodfellow}, \citenamefont {Bengio},\ and\ \citenamefont
  {Courville}}]{Goodfellow-et-al-2016}%
  \BibitemOpen
  \bibfield  {author} {\bibinfo {author} {\bibfnamefont {I.}~\bibnamefont
  {Goodfellow}}, \bibinfo {author} {\bibfnamefont {Y.}~\bibnamefont {Bengio}},
  \ and\ \bibinfo {author} {\bibfnamefont {A.}~\bibnamefont {Courville}},\
  }\href@noop {} {\emph {\bibinfo {title} {Deep Learning}}}\ (\bibinfo
  {publisher} {MIT Press},\ \bibinfo {year} {2016})\ \bibinfo {note}
  {\url{http://www.deeplearningbook.org}}\BibitemShut {NoStop}%
\bibitem [{\citenamefont {LeCun}\ \emph {et~al.}(2015)\citenamefont {LeCun},
  \citenamefont {Bengio},\ and\ \citenamefont {Hinton}}]{lecun2015deep}%
  \BibitemOpen
  \bibfield  {author} {\bibinfo {author} {\bibfnamefont {Y.}~\bibnamefont
  {LeCun}}, \bibinfo {author} {\bibfnamefont {Y.}~\bibnamefont {Bengio}}, \
  and\ \bibinfo {author} {\bibfnamefont {G.}~\bibnamefont {Hinton}},\
  }\href@noop {} {\bibfield  {journal} {\bibinfo  {journal} {nature}\ }\textbf
  {\bibinfo {volume} {521}},\ \bibinfo {pages} {436} (\bibinfo {year}
  {2015})}\BibitemShut {NoStop}%
\bibitem [{\citenamefont {Silver}\ \emph {et~al.}(2016)\citenamefont {Silver},
  \citenamefont {Huang}, \citenamefont {Maddison}, \citenamefont {Guez},
  \citenamefont {Sifre}, \citenamefont {Van Den~Driessche}, \citenamefont
  {Schrittwieser}, \citenamefont {Antonoglou}, \citenamefont {Panneershelvam},
  \citenamefont {Lanctot} \emph {et~al.}}]{silver2016mastering}%
  \BibitemOpen
  \bibfield  {author} {\bibinfo {author} {\bibfnamefont {D.}~\bibnamefont
  {Silver}}, \bibinfo {author} {\bibfnamefont {A.}~\bibnamefont {Huang}},
  \bibinfo {author} {\bibfnamefont {C.~J.}\ \bibnamefont {Maddison}}, \bibinfo
  {author} {\bibfnamefont {A.}~\bibnamefont {Guez}}, \bibinfo {author}
  {\bibfnamefont {L.}~\bibnamefont {Sifre}}, \bibinfo {author} {\bibfnamefont
  {G.}~\bibnamefont {Van Den~Driessche}}, \bibinfo {author} {\bibfnamefont
  {J.}~\bibnamefont {Schrittwieser}}, \bibinfo {author} {\bibfnamefont
  {I.}~\bibnamefont {Antonoglou}}, \bibinfo {author} {\bibfnamefont
  {V.}~\bibnamefont {Panneershelvam}}, \bibinfo {author} {\bibfnamefont
  {M.}~\bibnamefont {Lanctot}},  \emph {et~al.},\ }\href@noop {} {\bibfield
  {journal} {\bibinfo  {journal} {nature}\ }\textbf {\bibinfo {volume} {529}},\
  \bibinfo {pages} {484} (\bibinfo {year} {2016})}\BibitemShut {NoStop}%
\bibitem [{\citenamefont {Hezaveh}\ \emph {et~al.}(2017)\citenamefont
  {Hezaveh}, \citenamefont {Levasseur},\ and\ \citenamefont
  {Marshall}}]{hezaveh2017fast}%
  \BibitemOpen
  \bibfield  {author} {\bibinfo {author} {\bibfnamefont {Y.~D.}\ \bibnamefont
  {Hezaveh}}, \bibinfo {author} {\bibfnamefont {L.~P.}\ \bibnamefont
  {Levasseur}}, \ and\ \bibinfo {author} {\bibfnamefont {P.~J.}\ \bibnamefont
  {Marshall}},\ }\href@noop {} {\bibfield  {journal} {\bibinfo  {journal}
  {Nature}\ }\textbf {\bibinfo {volume} {548}},\ \bibinfo {pages} {555}
  (\bibinfo {year} {2017})}\BibitemShut {NoStop}%
\bibitem [{\citenamefont {Petrillo}\ \emph {et~al.}(2017)\citenamefont
  {Petrillo} \emph {et~al.}}]{Petrillo:2017njm}%
  \BibitemOpen
  \bibfield  {author} {\bibinfo {author} {\bibfnamefont {C.~E.}\ \bibnamefont
  {Petrillo}} \emph {et~al.},\ }\href {\doibase 10.1093/mnras/stx2052}
  {\bibfield  {journal} {\bibinfo  {journal} {Mon. Not. Roy. Astron. Soc.}\
  }\textbf {\bibinfo {volume} {472}},\ \bibinfo {pages} {1129} (\bibinfo {year}
  {2017})}\BibitemShut {NoStop}%
%%CITATION = ARXIV:1702.07675;%%
\bibitem [{\citenamefont {Sun}\ \emph {et~al.}(2018)\citenamefont {Sun},
  \citenamefont {Yi}, \citenamefont {Zhang}, \citenamefont {Shen},\ and\
  \citenamefont {Zhai}}]{sun2018deep}%
  \BibitemOpen
  \bibfield  {author} {\bibinfo {author} {\bibfnamefont {N.}~\bibnamefont
  {Sun}}, \bibinfo {author} {\bibfnamefont {J.}~\bibnamefont {Yi}}, \bibinfo
  {author} {\bibfnamefont {P.}~\bibnamefont {Zhang}}, \bibinfo {author}
  {\bibfnamefont {H.}~\bibnamefont {Shen}}, \ and\ \bibinfo {author}
  {\bibfnamefont {H.}~\bibnamefont {Zhai}},\ }\href@noop {} {\bibfield
  {journal} {\bibinfo  {journal} {Physical Review B}\ }\textbf {\bibinfo
  {volume} {98}},\ \bibinfo {pages} {085402} (\bibinfo {year}
  {2018})}\BibitemShut {NoStop}%
\bibitem [{\citenamefont {Carrasquilla}\ and\ \citenamefont
  {Melko}(2017)}]{carrasquilla2017machine}%
  \BibitemOpen
  \bibfield  {author} {\bibinfo {author} {\bibfnamefont {J.}~\bibnamefont
  {Carrasquilla}}\ and\ \bibinfo {author} {\bibfnamefont {R.~G.}\ \bibnamefont
  {Melko}},\ }\href@noop {} {\bibfield  {journal} {\bibinfo  {journal} {Nature
  Physics}\ }\textbf {\bibinfo {volume} {13}},\ \bibinfo {pages} {431}
  (\bibinfo {year} {2017})}\BibitemShut {NoStop}%
\bibitem [{\citenamefont {Van~Nieuwenburg}\ \emph {et~al.}(2017)\citenamefont
  {Van~Nieuwenburg}, \citenamefont {Liu},\ and\ \citenamefont
  {Huber}}]{van2017learning}%
  \BibitemOpen
  \bibfield  {author} {\bibinfo {author} {\bibfnamefont {E.~P.}\ \bibnamefont
  {Van~Nieuwenburg}}, \bibinfo {author} {\bibfnamefont {Y.-H.}\ \bibnamefont
  {Liu}}, \ and\ \bibinfo {author} {\bibfnamefont {S.~D.}\ \bibnamefont
  {Huber}},\ }\href@noop {} {\bibfield  {journal} {\bibinfo  {journal} {Nature
  Physics}\ }\textbf {\bibinfo {volume} {13}},\ \bibinfo {pages} {435}
  (\bibinfo {year} {2017})}\BibitemShut {NoStop}%
\bibitem [{\citenamefont {Broecker}\ \emph {et~al.}(2017)\citenamefont
  {Broecker}, \citenamefont {Carrasquilla}, \citenamefont {Melko},\ and\
  \citenamefont {Trebst}}]{broecker2017machine}%
  \BibitemOpen
  \bibfield  {author} {\bibinfo {author} {\bibfnamefont {P.}~\bibnamefont
  {Broecker}}, \bibinfo {author} {\bibfnamefont {J.}~\bibnamefont
  {Carrasquilla}}, \bibinfo {author} {\bibfnamefont {R.~G.}\ \bibnamefont
  {Melko}}, \ and\ \bibinfo {author} {\bibfnamefont {S.}~\bibnamefont
  {Trebst}},\ }\href@noop {} {\bibfield  {journal} {\bibinfo  {journal}
  {Scientific reports}\ }\textbf {\bibinfo {volume} {7}},\ \bibinfo {pages}
  {8823} (\bibinfo {year} {2017})}\BibitemShut {NoStop}%
\bibitem [{\citenamefont {Carleo}\ and\ \citenamefont
  {Troyer}(2017)}]{carleo2017solving}%
  \BibitemOpen
  \bibfield  {author} {\bibinfo {author} {\bibfnamefont {G.}~\bibnamefont
  {Carleo}}\ and\ \bibinfo {author} {\bibfnamefont {M.}~\bibnamefont
  {Troyer}},\ }\href@noop {} {\bibfield  {journal} {\bibinfo  {journal}
  {Science}\ }\textbf {\bibinfo {volume} {355}},\ \bibinfo {pages} {602}
  (\bibinfo {year} {2017})}\BibitemShut {NoStop}%
\bibitem [{\citenamefont {Gao}\ and\ \citenamefont
  {Duan}(2017)}]{gao2017efficient}%
  \BibitemOpen
  \bibfield  {author} {\bibinfo {author} {\bibfnamefont {X.}~\bibnamefont
  {Gao}}\ and\ \bibinfo {author} {\bibfnamefont {L.-M.}\ \bibnamefont {Duan}},\
  }\href@noop {} {\bibfield  {journal} {\bibinfo  {journal} {Nature
  communications}\ }\textbf {\bibinfo {volume} {8}},\ \bibinfo {pages} {662}
  (\bibinfo {year} {2017})}\BibitemShut {NoStop}%
\bibitem [{\citenamefont {Baldi}\ \emph {et~al.}(2014)\citenamefont {Baldi},
  \citenamefont {Sadowski},\ and\ \citenamefont
  {Whiteson}}]{baldi2014searching}%
  \BibitemOpen
  \bibfield  {author} {\bibinfo {author} {\bibfnamefont {P.}~\bibnamefont
  {Baldi}}, \bibinfo {author} {\bibfnamefont {P.}~\bibnamefont {Sadowski}}, \
  and\ \bibinfo {author} {\bibfnamefont {D.}~\bibnamefont {Whiteson}},\
  }\href@noop {} {\bibfield  {journal} {\bibinfo  {journal} {Nature
  communications}\ }\textbf {\bibinfo {volume} {5}},\ \bibinfo {pages} {4308}
  (\bibinfo {year} {2014})}\BibitemShut {NoStop}%
\bibitem [{\citenamefont {Baldi}\ \emph
  {et~al.}(2016{\natexlab{a}})\citenamefont {Baldi}, \citenamefont {Cranmer},
  \citenamefont {Faucett}, \citenamefont {Sadowski},\ and\ \citenamefont
  {Whiteson}}]{baldi2016parameterized}%
  \BibitemOpen
  \bibfield  {author} {\bibinfo {author} {\bibfnamefont {P.}~\bibnamefont
  {Baldi}}, \bibinfo {author} {\bibfnamefont {K.}~\bibnamefont {Cranmer}},
  \bibinfo {author} {\bibfnamefont {T.}~\bibnamefont {Faucett}}, \bibinfo
  {author} {\bibfnamefont {P.}~\bibnamefont {Sadowski}}, \ and\ \bibinfo
  {author} {\bibfnamefont {D.}~\bibnamefont {Whiteson}},\ }\href@noop {}
  {\bibfield  {journal} {\bibinfo  {journal} {The European Physical Journal C}\
  }\textbf {\bibinfo {volume} {76}},\ \bibinfo {pages} {235} (\bibinfo {year}
  {2016}{\natexlab{a}})}\BibitemShut {NoStop}%
\bibitem [{\citenamefont {Cogan}\ \emph {et~al.}(2015)\citenamefont {Cogan},
  \citenamefont {Kagan}, \citenamefont {Strauss},\ and\ \citenamefont
  {Schwarztman}}]{Cogan:2014oua}%
  \BibitemOpen
  \bibfield  {author} {\bibinfo {author} {\bibfnamefont {J.}~\bibnamefont
  {Cogan}}, \bibinfo {author} {\bibfnamefont {M.}~\bibnamefont {Kagan}},
  \bibinfo {author} {\bibfnamefont {E.}~\bibnamefont {Strauss}}, \ and\
  \bibinfo {author} {\bibfnamefont {A.}~\bibnamefont {Schwarztman}},\ }\href
  {\doibase 10.1007/JHEP02(2015)118} {\bibfield  {journal} {\bibinfo  {journal}
  {JHEP}\ }\textbf {\bibinfo {volume} {02}},\ \bibinfo {pages} {118} (\bibinfo
  {year} {2015})}\BibitemShut {NoStop}%
%%CITATION = ARXIV:1407.5675;%%
\bibitem [{\citenamefont {Baldi}\ \emph
  {et~al.}(2016{\natexlab{b}})\citenamefont {Baldi}, \citenamefont {Bauer},
  \citenamefont {Eng}, \citenamefont {Sadowski},\ and\ \citenamefont
  {Whiteson}}]{baldi2016jet}%
  \BibitemOpen
  \bibfield  {author} {\bibinfo {author} {\bibfnamefont {P.}~\bibnamefont
  {Baldi}}, \bibinfo {author} {\bibfnamefont {K.}~\bibnamefont {Bauer}},
  \bibinfo {author} {\bibfnamefont {C.}~\bibnamefont {Eng}}, \bibinfo {author}
  {\bibfnamefont {P.}~\bibnamefont {Sadowski}}, \ and\ \bibinfo {author}
  {\bibfnamefont {D.}~\bibnamefont {Whiteson}},\ }\href@noop {} {\bibfield
  {journal} {\bibinfo  {journal} {Physical Review D}\ }\textbf {\bibinfo
  {volume} {93}},\ \bibinfo {pages} {094034} (\bibinfo {year}
  {2016}{\natexlab{b}})}\BibitemShut {NoStop}%
\bibitem [{\citenamefont {Chakraborty}\ \emph {et~al.}(2019)\citenamefont
  {Chakraborty}, \citenamefont {Lim},\ and\ \citenamefont
  {Nojiri}}]{chakraborty2019interpretable}%
  \BibitemOpen
  \bibfield  {author} {\bibinfo {author} {\bibfnamefont {A.}~\bibnamefont
  {Chakraborty}}, \bibinfo {author} {\bibfnamefont {S.~H.}\ \bibnamefont
  {Lim}}, \ and\ \bibinfo {author} {\bibfnamefont {M.~M.}\ \bibnamefont
  {Nojiri}},\ }\href@noop {} {\bibfield  {journal} {\bibinfo  {journal}
  {Journal of High Energy Physics}\ }\textbf {\bibinfo {volume} {2019}},\
  \bibinfo {pages} {135} (\bibinfo {year} {2019})}\BibitemShut {NoStop}%
\bibitem [{\citenamefont {Pang}\ \emph
  {et~al.}(2018{\natexlab{a}})\citenamefont {Pang}, \citenamefont {Zhou},
  \citenamefont {Su}, \citenamefont {Petersen}, \citenamefont {St{\"o}cker},\
  and\ \citenamefont {Wang}}]{pang2018equation}%
  \BibitemOpen
  \bibfield  {author} {\bibinfo {author} {\bibfnamefont {L.-G.}\ \bibnamefont
  {Pang}}, \bibinfo {author} {\bibfnamefont {K.}~\bibnamefont {Zhou}}, \bibinfo
  {author} {\bibfnamefont {N.}~\bibnamefont {Su}}, \bibinfo {author}
  {\bibfnamefont {H.}~\bibnamefont {Petersen}}, \bibinfo {author}
  {\bibfnamefont {H.}~\bibnamefont {St{\"o}cker}}, \ and\ \bibinfo {author}
  {\bibfnamefont {X.-N.}\ \bibnamefont {Wang}},\ }\href@noop {} {\bibfield
  {journal} {\bibinfo  {journal} {Nature communications}\ }\textbf {\bibinfo
  {volume} {9}},\ \bibinfo {pages} {210} (\bibinfo {year}
  {2018}{\natexlab{a}})}\BibitemShut {NoStop}%
\bibitem [{\citenamefont {Chien}(2019)}]{chien2019probing}%
  \BibitemOpen
  \bibfield  {author} {\bibinfo {author} {\bibfnamefont {Y.-T.}\ \bibnamefont
  {Chien}},\ }\href@noop {} {\bibfield  {journal} {\bibinfo  {journal} {Nuclear
  Physics A}\ }\textbf {\bibinfo {volume} {982}},\ \bibinfo {pages} {619}
  (\bibinfo {year} {2019})}\BibitemShut {NoStop}%
\bibitem [{\citenamefont {Steinheimer}\ \emph {et~al.}(2019)\citenamefont
  {Steinheimer}, \citenamefont {Pang}, \citenamefont {Zhou}, \citenamefont
  {Koch}, \citenamefont {Randrup},\ and\ \citenamefont
  {Stoecker}}]{steinheimer2019machine}%
  \BibitemOpen
  \bibfield  {author} {\bibinfo {author} {\bibfnamefont {J.}~\bibnamefont
  {Steinheimer}}, \bibinfo {author} {\bibfnamefont {L.-G.}\ \bibnamefont
  {Pang}}, \bibinfo {author} {\bibfnamefont {K.}~\bibnamefont {Zhou}}, \bibinfo
  {author} {\bibfnamefont {V.}~\bibnamefont {Koch}}, \bibinfo {author}
  {\bibfnamefont {J.}~\bibnamefont {Randrup}}, \ and\ \bibinfo {author}
  {\bibfnamefont {H.}~\bibnamefont {Stoecker}},\ }\href@noop {} {\bibfield
  {journal} {\bibinfo  {journal} {Journal of High Energy Physics}\ }\textbf
  {\bibinfo {volume} {2019}},\ \bibinfo {pages} {122} (\bibinfo {year}
  {2019})}\BibitemShut {NoStop}%
\bibitem [{\citenamefont {Pang}\ \emph {et~al.}(2019)\citenamefont {Pang},
  \citenamefont {Zhou},\ and\ \citenamefont {Wang}}]{pang2019interpretable}%
  \BibitemOpen
  \bibfield  {author} {\bibinfo {author} {\bibfnamefont {L.-G.}\ \bibnamefont
  {Pang}}, \bibinfo {author} {\bibfnamefont {K.}~\bibnamefont {Zhou}}, \ and\
  \bibinfo {author} {\bibfnamefont {X.-N.}\ \bibnamefont {Wang}},\ }\href@noop
  {} {\bibfield  {journal} {\bibinfo  {journal} {arXiv preprint
  arXiv:1906.06429}\ } (\bibinfo {year} {2019})}\BibitemShut {NoStop}%
\bibitem [{\citenamefont {Bernhard}\ \emph {et~al.}(2016)\citenamefont
  {Bernhard}, \citenamefont {Moreland}, \citenamefont {Bass}, \citenamefont
  {Liu},\ and\ \citenamefont {Heinz}}]{Bernhard:2016tnd}%
  \BibitemOpen
  \bibfield  {author} {\bibinfo {author} {\bibfnamefont {J.~E.}\ \bibnamefont
  {Bernhard}}, \bibinfo {author} {\bibfnamefont {J.~S.}\ \bibnamefont
  {Moreland}}, \bibinfo {author} {\bibfnamefont {S.~A.}\ \bibnamefont {Bass}},
  \bibinfo {author} {\bibfnamefont {J.}~\bibnamefont {Liu}}, \ and\ \bibinfo
  {author} {\bibfnamefont {U.}~\bibnamefont {Heinz}},\ }\href {\doibase
  10.1103/PhysRevC.94.024907} {\bibfield  {journal} {\bibinfo  {journal} {Phys.
  Rev. C}\ }\textbf {\bibinfo {volume} {94}},\ \bibinfo {pages} {024907}
  (\bibinfo {year} {2016})}\BibitemShut {NoStop}%
\bibitem [{\citenamefont {Bernhard}\ \emph {et~al.}(2019)\citenamefont
  {Bernhard}, \citenamefont {Moreland},\ and\ \citenamefont
  {Bass}}]{Bernhard:2019bmu}%
  \BibitemOpen
  \bibfield  {author} {\bibinfo {author} {\bibfnamefont {J.~E.}\ \bibnamefont
  {Bernhard}}, \bibinfo {author} {\bibfnamefont {J.~S.}\ \bibnamefont
  {Moreland}}, \ and\ \bibinfo {author} {\bibfnamefont {S.~A.}\ \bibnamefont
  {Bass}},\ }\href {\doibase 10.1038/s41567-019-0611-8} {\bibfield  {journal}
  {\bibinfo  {journal} {Nature Phys.}\ }\textbf {\bibinfo {volume} {15}},\
  \bibinfo {pages} {1113} (\bibinfo {year} {2019})}\BibitemShut {NoStop}%
\bibitem [{\citenamefont {Everett}\ \emph
  {et~al.}(2020{\natexlab{a}})\citenamefont {Everett} \emph
  {et~al.}}]{Everett:2020yty}%
  \BibitemOpen
  \bibfield  {author} {\bibinfo {author} {\bibfnamefont {D.}~\bibnamefont
  {Everett}} \emph {et~al.} (\bibinfo {collaboration} {JETSCAPE}),\ }\href@noop
  {} {\  (\bibinfo {year} {2020}{\natexlab{a}})},\ \Eprint
  {http://arxiv.org/abs/2010.03928} {arXiv:2010.03928 [hep-ph]} \BibitemShut
  {NoStop}%
\bibitem [{\citenamefont {Everett}\ \emph
  {et~al.}(2020{\natexlab{b}})\citenamefont {Everett} \emph
  {et~al.}}]{Everett:2020xug}%
  \BibitemOpen
  \bibfield  {author} {\bibinfo {author} {\bibfnamefont {D.}~\bibnamefont
  {Everett}} \emph {et~al.} (\bibinfo {collaboration} {JETSCAPE}),\ }\href@noop
  {} {\  (\bibinfo {year} {2020}{\natexlab{b}})},\ \Eprint
  {http://arxiv.org/abs/2011.01430} {arXiv:2011.01430 [hep-ph]} \BibitemShut
  {NoStop}%
\bibitem [{\citenamefont {Zhou}\ \emph {et~al.}(2019)\citenamefont {Zhou},
  \citenamefont {Endr{\H{o}}di}, \citenamefont {Pang},\ and\ \citenamefont
  {St{\"o}cker}}]{zhou2019regressive}%
  \BibitemOpen
  \bibfield  {author} {\bibinfo {author} {\bibfnamefont {K.}~\bibnamefont
  {Zhou}}, \bibinfo {author} {\bibfnamefont {G.}~\bibnamefont {Endr{\H{o}}di}},
  \bibinfo {author} {\bibfnamefont {L.-G.}\ \bibnamefont {Pang}}, \ and\
  \bibinfo {author} {\bibfnamefont {H.}~\bibnamefont {St{\"o}cker}},\
  }\href@noop {} {\bibfield  {journal} {\bibinfo  {journal} {Physical Review
  D}\ }\textbf {\bibinfo {volume} {100}},\ \bibinfo {pages} {011501} (\bibinfo
  {year} {2019})}\BibitemShut {NoStop}%
\bibitem [{\citenamefont {Bhalerao}\ \emph {et~al.}(2015)\citenamefont
  {Bhalerao}, \citenamefont {Ollitrault}, \citenamefont {Pal},\ and\
  \citenamefont {Teaney}}]{PhysRevLett.114.152301}%
  \BibitemOpen
  \bibfield  {author} {\bibinfo {author} {\bibfnamefont {R.~S.}\ \bibnamefont
  {Bhalerao}}, \bibinfo {author} {\bibfnamefont {J.-Y.}\ \bibnamefont
  {Ollitrault}}, \bibinfo {author} {\bibfnamefont {S.}~\bibnamefont {Pal}}, \
  and\ \bibinfo {author} {\bibfnamefont {D.}~\bibnamefont {Teaney}},\ }\href
  {\doibase 10.1103/PhysRevLett.114.152301} {\bibfield  {journal} {\bibinfo
  {journal} {Phys. Rev. Lett.}\ }\textbf {\bibinfo {volume} {114}},\ \bibinfo
  {pages} {152301} (\bibinfo {year} {2015})}\BibitemShut {NoStop}%
\bibitem [{\citenamefont {Mazeliauskas}\ and\ \citenamefont
  {Teaney}(2015)}]{mazeliauskas2015subleading}%
  \BibitemOpen
  \bibfield  {author} {\bibinfo {author} {\bibfnamefont {A.}~\bibnamefont
  {Mazeliauskas}}\ and\ \bibinfo {author} {\bibfnamefont {D.}~\bibnamefont
  {Teaney}},\ }\href@noop {} {\bibfield  {journal} {\bibinfo  {journal}
  {Physical Review C}\ }\textbf {\bibinfo {volume} {91}},\ \bibinfo {pages}
  {044902} (\bibinfo {year} {2015})}\BibitemShut {NoStop}%
\bibitem [{\citenamefont {Sirunyan}\ \emph {et~al.}(2017)\citenamefont
  {Sirunyan}, \citenamefont {Tumasyan}, \citenamefont {Adam}, \citenamefont
  {Ambrogi}, \citenamefont {Asilar}, \citenamefont {Bergauer}, \citenamefont
  {Brandstetter}, \citenamefont {Brondolin}, \citenamefont {Dragicevic},
  \citenamefont {Er{\"o}} \emph {et~al.}}]{sirunyan2017principal}%
  \BibitemOpen
  \bibfield  {author} {\bibinfo {author} {\bibfnamefont {A.~M.}\ \bibnamefont
  {Sirunyan}}, \bibinfo {author} {\bibfnamefont {A.}~\bibnamefont {Tumasyan}},
  \bibinfo {author} {\bibfnamefont {W.}~\bibnamefont {Adam}}, \bibinfo {author}
  {\bibfnamefont {F.}~\bibnamefont {Ambrogi}}, \bibinfo {author} {\bibfnamefont
  {E.}~\bibnamefont {Asilar}}, \bibinfo {author} {\bibfnamefont
  {T.}~\bibnamefont {Bergauer}}, \bibinfo {author} {\bibfnamefont
  {J.}~\bibnamefont {Brandstetter}}, \bibinfo {author} {\bibfnamefont
  {E.}~\bibnamefont {Brondolin}}, \bibinfo {author} {\bibfnamefont
  {M.}~\bibnamefont {Dragicevic}}, \bibinfo {author} {\bibfnamefont
  {J.}~\bibnamefont {Er{\"o}}},  \emph {et~al.},\ }\href@noop {} {\bibfield
  {journal} {\bibinfo  {journal} {Physical Review C}\ }\textbf {\bibinfo
  {volume} {96}},\ \bibinfo {pages} {064902} (\bibinfo {year}
  {2017})}\BibitemShut {NoStop}%
\bibitem [{\citenamefont {Liu}\ \emph {et~al.}(2019)\citenamefont {Liu},
  \citenamefont {Zhao},\ and\ \citenamefont {Song}}]{liu2019principal}%
  \BibitemOpen
  \bibfield  {author} {\bibinfo {author} {\bibfnamefont {Z.}~\bibnamefont
  {Liu}}, \bibinfo {author} {\bibfnamefont {W.}~\bibnamefont {Zhao}}, \ and\
  \bibinfo {author} {\bibfnamefont {H.}~\bibnamefont {Song}},\ }\href@noop {}
  {\bibfield  {journal} {\bibinfo  {journal} {The European Physical Journal C}\
  }\textbf {\bibinfo {volume} {79}},\ \bibinfo {pages} {870} (\bibinfo {year}
  {2019})}\BibitemShut {NoStop}%
\bibitem [{\citenamefont {Liu}\ \emph {et~al.}(2020)\citenamefont {Liu},
  \citenamefont {Behera}, \citenamefont {Song},\ and\ \citenamefont
  {Jia}}]{Liu:2020ely}%
  \BibitemOpen
  \bibfield  {author} {\bibinfo {author} {\bibfnamefont {Z.}~\bibnamefont
  {Liu}}, \bibinfo {author} {\bibfnamefont {A.}~\bibnamefont {Behera}},
  \bibinfo {author} {\bibfnamefont {H.}~\bibnamefont {Song}}, \ and\ \bibinfo
  {author} {\bibfnamefont {J.}~\bibnamefont {Jia}},\ }\href {\doibase
  10.1103/PhysRevC.102.024911} {\bibfield  {journal} {\bibinfo  {journal}
  {Phys. Rev. C}\ }\textbf {\bibinfo {volume} {102}},\ \bibinfo {pages}
  {024911} (\bibinfo {year} {2020})}\BibitemShut {NoStop}%
\bibitem [{\citenamefont {Gardim}\ \emph {et~al.}(2019)\citenamefont {Gardim},
  \citenamefont {Grassi}, \citenamefont {Ishida}, \citenamefont {Luzum},\ and\
  \citenamefont {Ollitrault}}]{PhysRevC.100.054905}%
  \BibitemOpen
  \bibfield  {author} {\bibinfo {author} {\bibfnamefont {F.~G.}\ \bibnamefont
  {Gardim}}, \bibinfo {author} {\bibfnamefont {F.}~\bibnamefont {Grassi}},
  \bibinfo {author} {\bibfnamefont {P.}~\bibnamefont {Ishida}}, \bibinfo
  {author} {\bibfnamefont {M.}~\bibnamefont {Luzum}}, \ and\ \bibinfo {author}
  {\bibfnamefont {J.-Y.}\ \bibnamefont {Ollitrault}},\ }\href {\doibase
  10.1103/PhysRevC.100.054905} {\bibfield  {journal} {\bibinfo  {journal}
  {Phys. Rev. C}\ }\textbf {\bibinfo {volume} {100}},\ \bibinfo {pages}
  {054905} (\bibinfo {year} {2019})}\BibitemShut {NoStop}%
\bibitem [{\citenamefont {Altsybeev}(2020)}]{Altsybeev_2020}%
  \BibitemOpen
  \bibfield  {author} {\bibinfo {author} {\bibfnamefont {I.}~\bibnamefont
  {Altsybeev}},\ }\href {\doibase 10.1088/1742-6596/1602/1/012004} {\bibfield
  {journal} {\bibinfo  {journal} {Journal of Physics: Conference Series}\
  }\textbf {\bibinfo {volume} {1602}},\ \bibinfo {pages} {012004} (\bibinfo
  {year} {2020})}\BibitemShut {NoStop}%
\bibitem [{\citenamefont {Landau}\ and\ \citenamefont
  {Lifshitz}(2013)}]{landau2013course}%
  \BibitemOpen
  \bibfield  {author} {\bibinfo {author} {\bibfnamefont {L.~D.}\ \bibnamefont
  {Landau}}\ and\ \bibinfo {author} {\bibfnamefont {E.~M.}\ \bibnamefont
  {Lifshitz}},\ }\href@noop {} {}\ (\bibinfo  {publisher} {Elsevier},\ \bibinfo
  {year} {2013})\BibitemShut {NoStop}%
\bibitem [{\citenamefont {Gyulassy}\ and\ \citenamefont
  {McLerran}(2005)}]{gyulassy2005new}%
  \BibitemOpen
  \bibfield  {author} {\bibinfo {author} {\bibfnamefont {M.}~\bibnamefont
  {Gyulassy}}\ and\ \bibinfo {author} {\bibfnamefont {L.}~\bibnamefont
  {McLerran}},\ }\href@noop {} {\bibfield  {journal} {\bibinfo  {journal}
  {Nuclear Physics A}\ }\textbf {\bibinfo {volume} {750}},\ \bibinfo {pages}
  {30} (\bibinfo {year} {2005})}\BibitemShut {NoStop}%
\bibitem [{\citenamefont {M{\"u}ller}\ and\ \citenamefont
  {Nagle}(2006)}]{muller2006results}%
  \BibitemOpen
  \bibfield  {author} {\bibinfo {author} {\bibfnamefont {B.}~\bibnamefont
  {M{\"u}ller}}\ and\ \bibinfo {author} {\bibfnamefont {J.~L.}\ \bibnamefont
  {Nagle}},\ }\href@noop {} {\bibfield  {journal} {\bibinfo  {journal} {Annu.
  Rev. Nucl. Part. Sci.}\ }\textbf {\bibinfo {volume} {56}},\ \bibinfo {pages}
  {93} (\bibinfo {year} {2006})}\BibitemShut {NoStop}%
\bibitem [{\citenamefont {Huovinen}(2004)}]{huovinen2004hydrodynamical}%
  \BibitemOpen
  \bibfield  {author} {\bibinfo {author} {\bibfnamefont {P.}~\bibnamefont
  {Huovinen}},\ }in\ \href@noop {} {\emph {\bibinfo {booktitle} {Quark--Gluon
  Plasma 3}}}\ (\bibinfo  {publisher} {World Scientific},\ \bibinfo {year}
  {2004})\ pp.\ \bibinfo {pages} {600--633}\BibitemShut {NoStop}%
\bibitem [{\citenamefont {Kolb}\ and\ \citenamefont
  {Heinz}(2004)}]{kolb2004hydrodynamic}%
  \BibitemOpen
  \bibfield  {author} {\bibinfo {author} {\bibfnamefont {P.~F.}\ \bibnamefont
  {Kolb}}\ and\ \bibinfo {author} {\bibfnamefont {U.}~\bibnamefont {Heinz}},\
  }in\ \href@noop {} {\emph {\bibinfo {booktitle} {Quark--Gluon Plasma 3}}}\
  (\bibinfo  {publisher} {World Scientific},\ \bibinfo {year} {2004})\ pp.\
  \bibinfo {pages} {634--714}\BibitemShut {NoStop}%
\bibitem [{\citenamefont {Heinz}\ and\ \citenamefont
  {Snellings}(2013)}]{heinz2013collective}%
  \BibitemOpen
  \bibfield  {author} {\bibinfo {author} {\bibfnamefont {U.}~\bibnamefont
  {Heinz}}\ and\ \bibinfo {author} {\bibfnamefont {R.}~\bibnamefont
  {Snellings}},\ }\href@noop {} {\bibfield  {journal} {\bibinfo  {journal}
  {Annual Review of Nuclear and Particle Science}\ }\textbf {\bibinfo {volume}
  {63}},\ \bibinfo {pages} {123} (\bibinfo {year} {2013})}\BibitemShut
  {NoStop}%
\bibitem [{\citenamefont {Gale}\ \emph
  {et~al.}(2013{\natexlab{a}})\citenamefont {Gale}, \citenamefont {Jeon},\ and\
  \citenamefont {Schenke}}]{gale2013hydrodynamic}%
  \BibitemOpen
  \bibfield  {author} {\bibinfo {author} {\bibfnamefont {C.}~\bibnamefont
  {Gale}}, \bibinfo {author} {\bibfnamefont {S.}~\bibnamefont {Jeon}}, \ and\
  \bibinfo {author} {\bibfnamefont {B.}~\bibnamefont {Schenke}},\ }\href@noop
  {} {\bibfield  {journal} {\bibinfo  {journal} {International Journal of
  Modern Physics A}\ }\textbf {\bibinfo {volume} {28}},\ \bibinfo {pages}
  {1340011} (\bibinfo {year} {2013}{\natexlab{a}})}\BibitemShut {NoStop}%
\bibitem [{\citenamefont {Song}\ \emph {et~al.}(2017)\citenamefont {Song},
  \citenamefont {Zhou},\ and\ \citenamefont
  {Gajdo{\v{s}}ov{\'a}}}]{song2017collective}%
  \BibitemOpen
  \bibfield  {author} {\bibinfo {author} {\bibfnamefont {H.}~\bibnamefont
  {Song}}, \bibinfo {author} {\bibfnamefont {Y.}~\bibnamefont {Zhou}}, \ and\
  \bibinfo {author} {\bibfnamefont {K.}~\bibnamefont {Gajdo{\v{s}}ov{\'a}}},\
  }\href@noop {} {\bibfield  {journal} {\bibinfo  {journal} {Nuclear Science
  and Techniques}\ }\textbf {\bibinfo {volume} {28}},\ \bibinfo {pages} {99}
  (\bibinfo {year} {2017})}\BibitemShut {NoStop}%
\bibitem [{\citenamefont {Muller}\ \emph {et~al.}()\citenamefont {Muller},
  \citenamefont {Schukraft},\ and\ \citenamefont {Wyslouch}}]{muller3233first}%
  \BibitemOpen
  \bibfield  {author} {\bibinfo {author} {\bibfnamefont {B.}~\bibnamefont
  {Muller}}, \bibinfo {author} {\bibfnamefont {J.}~\bibnamefont {Schukraft}}, \
  and\ \bibinfo {author} {\bibfnamefont {B.}~\bibnamefont {Wyslouch}},\
  }\href@noop {} {\bibfield  {journal} {\bibinfo  {journal} {arXiv preprint
  arXiv:1202.3233}\ }\textbf {\bibinfo {volume} {1}}}\BibitemShut {NoStop}%
\bibitem [{\citenamefont {Song}\ \emph {et~al.}(2011)\citenamefont {Song},
  \citenamefont {Bass}, \citenamefont {Heinz}, \citenamefont {Hirano},\ and\
  \citenamefont {Shen}}]{Song:2010mg}%
  \BibitemOpen
  \bibfield  {author} {\bibinfo {author} {\bibfnamefont {H.}~\bibnamefont
  {Song}}, \bibinfo {author} {\bibfnamefont {S.~A.}\ \bibnamefont {Bass}},
  \bibinfo {author} {\bibfnamefont {U.}~\bibnamefont {Heinz}}, \bibinfo
  {author} {\bibfnamefont {T.}~\bibnamefont {Hirano}}, \ and\ \bibinfo {author}
  {\bibfnamefont {C.}~\bibnamefont {Shen}},\ }\href {\doibase
  10.1103/PhysRevLett.106.192301} {\bibfield  {journal} {\bibinfo  {journal}
  {Phys. Rev. Lett.}\ }\textbf {\bibinfo {volume} {106}},\ \bibinfo {pages}
  {192301} (\bibinfo {year} {2011})}\BibitemShut {NoStop}%
\bibitem [{\citenamefont {Schenke}\ \emph {et~al.}(2011)\citenamefont
  {Schenke}, \citenamefont {Jeon},\ and\ \citenamefont
  {Gale}}]{Schenke:2010rr}%
  \BibitemOpen
  \bibfield  {author} {\bibinfo {author} {\bibfnamefont {B.}~\bibnamefont
  {Schenke}}, \bibinfo {author} {\bibfnamefont {S.}~\bibnamefont {Jeon}}, \
  and\ \bibinfo {author} {\bibfnamefont {C.}~\bibnamefont {Gale}},\ }\href
  {\doibase 10.1103/PhysRevLett.106.042301} {\bibfield  {journal} {\bibinfo
  {journal} {Phys. Rev. Lett.}\ }\textbf {\bibinfo {volume} {106}},\ \bibinfo
  {pages} {042301} (\bibinfo {year} {2011})}\BibitemShut {NoStop}%
\bibitem [{\citenamefont {Gale}\ \emph
  {et~al.}(2013{\natexlab{b}})\citenamefont {Gale}, \citenamefont {Jeon},
  \citenamefont {Schenke}, \citenamefont {Tribedy},\ and\ \citenamefont
  {Venugopalan}}]{Gale:2012rq}%
  \BibitemOpen
  \bibfield  {author} {\bibinfo {author} {\bibfnamefont {C.}~\bibnamefont
  {Gale}}, \bibinfo {author} {\bibfnamefont {S.}~\bibnamefont {Jeon}}, \bibinfo
  {author} {\bibfnamefont {B.}~\bibnamefont {Schenke}}, \bibinfo {author}
  {\bibfnamefont {P.}~\bibnamefont {Tribedy}}, \ and\ \bibinfo {author}
  {\bibfnamefont {R.}~\bibnamefont {Venugopalan}},\ }\href {\doibase
  10.1103/PhysRevLett.110.012302} {\bibfield  {journal} {\bibinfo  {journal}
  {Phys. Rev. Lett.}\ }\textbf {\bibinfo {volume} {110}},\ \bibinfo {pages}
  {012302} (\bibinfo {year} {2013}{\natexlab{b}})}\BibitemShut {NoStop}%
\bibitem [{\citenamefont {Zhu}\ \emph {et~al.}(2017)\citenamefont {Zhu},
  \citenamefont {Zhou}, \citenamefont {Xu},\ and\ \citenamefont
  {Song}}]{Zhu:2016puf}%
  \BibitemOpen
  \bibfield  {author} {\bibinfo {author} {\bibfnamefont {X.}~\bibnamefont
  {Zhu}}, \bibinfo {author} {\bibfnamefont {Y.}~\bibnamefont {Zhou}}, \bibinfo
  {author} {\bibfnamefont {H.}~\bibnamefont {Xu}}, \ and\ \bibinfo {author}
  {\bibfnamefont {H.}~\bibnamefont {Song}},\ }\href {\doibase
  10.1103/PhysRevC.95.044902} {\bibfield  {journal} {\bibinfo  {journal} {Phys.
  Rev. C}\ }\textbf {\bibinfo {volume} {95}},\ \bibinfo {pages} {044902}
  (\bibinfo {year} {2017})}\BibitemShut {NoStop}%
\bibitem [{\citenamefont {Zhao}\ \emph {et~al.}(2017)\citenamefont {Zhao},
  \citenamefont {Xu},\ and\ \citenamefont {Song}}]{Zhao:2017yhj}%
  \BibitemOpen
  \bibfield  {author} {\bibinfo {author} {\bibfnamefont {W.}~\bibnamefont
  {Zhao}}, \bibinfo {author} {\bibfnamefont {H.-j.}\ \bibnamefont {Xu}}, \ and\
  \bibinfo {author} {\bibfnamefont {H.}~\bibnamefont {Song}},\ }\href {\doibase
  10.1140/epjc/s10052-017-5186-x} {\bibfield  {journal} {\bibinfo  {journal}
  {Eur. Phys. J. C}\ }\textbf {\bibinfo {volume} {77}},\ \bibinfo {pages} {645}
  (\bibinfo {year} {2017})}\BibitemShut {NoStop}%
\bibitem [{\citenamefont {Shen}\ \emph {et~al.}(2016)\citenamefont {Shen},
  \citenamefont {Qiu}, \citenamefont {Song}, \citenamefont {Bernhard},
  \citenamefont {Bass},\ and\ \citenamefont {Heinz}}]{shen2016iebe}%
  \BibitemOpen
  \bibfield  {author} {\bibinfo {author} {\bibfnamefont {C.}~\bibnamefont
  {Shen}}, \bibinfo {author} {\bibfnamefont {Z.}~\bibnamefont {Qiu}}, \bibinfo
  {author} {\bibfnamefont {H.}~\bibnamefont {Song}}, \bibinfo {author}
  {\bibfnamefont {J.}~\bibnamefont {Bernhard}}, \bibinfo {author}
  {\bibfnamefont {S.}~\bibnamefont {Bass}}, \ and\ \bibinfo {author}
  {\bibfnamefont {U.}~\bibnamefont {Heinz}},\ }\href@noop {} {\bibfield
  {journal} {\bibinfo  {journal} {Computer Physics Communications}\ }\textbf
  {\bibinfo {volume} {199}},\ \bibinfo {pages} {61} (\bibinfo {year}
  {2016})}\BibitemShut {NoStop}%
\bibitem [{\citenamefont {Pratt}\ \emph {et~al.}(2015)\citenamefont {Pratt},
  \citenamefont {Sangaline}, \citenamefont {Sorensen},\ and\ \citenamefont
  {Wang}}]{Pratt:2015zsa}%
  \BibitemOpen
  \bibfield  {author} {\bibinfo {author} {\bibfnamefont {S.}~\bibnamefont
  {Pratt}}, \bibinfo {author} {\bibfnamefont {E.}~\bibnamefont {Sangaline}},
  \bibinfo {author} {\bibfnamefont {P.}~\bibnamefont {Sorensen}}, \ and\
  \bibinfo {author} {\bibfnamefont {H.}~\bibnamefont {Wang}},\ }\href {\doibase
  10.1103/PhysRevLett.114.202301} {\bibfield  {journal} {\bibinfo  {journal}
  {Phys. Rev. Lett.}\ }\textbf {\bibinfo {volume} {114}},\ \bibinfo {pages}
  {202301} (\bibinfo {year} {2015})}\BibitemShut {NoStop}%
\bibitem [{\citenamefont {Song}\ and\ \citenamefont
  {Heinz}(2008)}]{song2008causal}%
  \BibitemOpen
  \bibfield  {author} {\bibinfo {author} {\bibfnamefont {H.}~\bibnamefont
  {Song}}\ and\ \bibinfo {author} {\bibfnamefont {U.}~\bibnamefont {Heinz}},\
  }\href@noop {} {\bibfield  {journal} {\bibinfo  {journal} {Physical Review
  C}\ }\textbf {\bibinfo {volume} {77}},\ \bibinfo {pages} {064901} (\bibinfo
  {year} {2008})}\BibitemShut {NoStop}%
\bibitem [{Note1()}]{Note1}%
  \BibitemOpen
  \bibinfo {note} {In $(\tau ,x,y,\eta )$ coordinate ($\tau =\protect \sqrt
  {t^2-z^2}$ and $\eta =\protect \frac {1}{2}\protect \mathrm {ln}\protect
  \frac {t+z}{t-z}$), the energy density and pressure from 2+1-d hydrodynamics
  are longitudinal boost-invariant without a dependence on $\eta $, $e$=$e(\tau
  ,x,y)$ and $p$=$p(\tau ,x,y)$. Correspondingly, the four flow velocity are
  expressed as $\gamma _\bot (1, v^x(\tau ,x, y), v^y(\tau ,x, y), 0)$ with
  $v^\eta =0$~\cite
  {huovinen2004hydrodynamical,kolb2004hydrodynamic}.}\BibitemShut {Stop}%
\bibitem [{\citenamefont {Miller}\ \emph {et~al.}(2007)\citenamefont {Miller},
  \citenamefont {Reygers}, \citenamefont {Sanders},\ and\ \citenamefont
  {Steinberg}}]{miller2007glauber}%
  \BibitemOpen
  \bibfield  {author} {\bibinfo {author} {\bibfnamefont {M.~L.}\ \bibnamefont
  {Miller}}, \bibinfo {author} {\bibfnamefont {K.}~\bibnamefont {Reygers}},
  \bibinfo {author} {\bibfnamefont {S.~J.}\ \bibnamefont {Sanders}}, \ and\
  \bibinfo {author} {\bibfnamefont {P.}~\bibnamefont {Steinberg}},\ }\href@noop
  {} {\bibfield  {journal} {\bibinfo  {journal} {Annu. Rev. Nucl. Part. Sci.}\
  }\textbf {\bibinfo {volume} {57}},\ \bibinfo {pages} {205} (\bibinfo {year}
  {2007})}\BibitemShut {NoStop}%
\bibitem [{\citenamefont {Hirano}\ and\ \citenamefont
  {Nara}(2009)}]{hirano2009eccentricity}%
  \BibitemOpen
  \bibfield  {author} {\bibinfo {author} {\bibfnamefont {T.}~\bibnamefont
  {Hirano}}\ and\ \bibinfo {author} {\bibfnamefont {Y.}~\bibnamefont {Nara}},\
  }\href@noop {} {\bibfield  {journal} {\bibinfo  {journal} {Physical Review
  C}\ }\textbf {\bibinfo {volume} {79}},\ \bibinfo {pages} {064904} (\bibinfo
  {year} {2009})}\BibitemShut {NoStop}%
\bibitem [{\citenamefont {Drescher}\ and\ \citenamefont
  {Nara}(2007{\natexlab{a}})}]{drescher2007effects}%
  \BibitemOpen
  \bibfield  {author} {\bibinfo {author} {\bibfnamefont {H.-J.}\ \bibnamefont
  {Drescher}}\ and\ \bibinfo {author} {\bibfnamefont {Y.}~\bibnamefont
  {Nara}},\ }\href@noop {} {\bibfield  {journal} {\bibinfo  {journal} {Physical
  Review C}\ }\textbf {\bibinfo {volume} {75}},\ \bibinfo {pages} {034905}
  (\bibinfo {year} {2007}{\natexlab{a}})}\BibitemShut {NoStop}%
\bibitem [{\citenamefont {Drescher}\ and\ \citenamefont
  {Nara}(2007{\natexlab{b}})}]{drescher2007eccentricity}%
  \BibitemOpen
  \bibfield  {author} {\bibinfo {author} {\bibfnamefont {H.-J.}\ \bibnamefont
  {Drescher}}\ and\ \bibinfo {author} {\bibfnamefont {Y.}~\bibnamefont
  {Nara}},\ }\href@noop {} {\bibfield  {journal} {\bibinfo  {journal} {Physical
  Review C}\ }\textbf {\bibinfo {volume} {76}},\ \bibinfo {pages} {041903}
  (\bibinfo {year} {2007}{\natexlab{b}})}\BibitemShut {NoStop}%
\bibitem [{\citenamefont {Pang}\ \emph {et~al.}(2012)\citenamefont {Pang},
  \citenamefont {Wang},\ and\ \citenamefont {Wang}}]{pang2012effects}%
  \BibitemOpen
  \bibfield  {author} {\bibinfo {author} {\bibfnamefont {L.}~\bibnamefont
  {Pang}}, \bibinfo {author} {\bibfnamefont {Q.}~\bibnamefont {Wang}}, \ and\
  \bibinfo {author} {\bibfnamefont {X.-N.}\ \bibnamefont {Wang}},\ }\href@noop
  {} {\bibfield  {journal} {\bibinfo  {journal} {Physical Review C}\ }\textbf
  {\bibinfo {volume} {86}},\ \bibinfo {pages} {024911} (\bibinfo {year}
  {2012})}\BibitemShut {NoStop}%
\bibitem [{\citenamefont {Xu}\ \emph {et~al.}(2016)\citenamefont {Xu},
  \citenamefont {Li},\ and\ \citenamefont {Song}}]{xu2016high}%
  \BibitemOpen
  \bibfield  {author} {\bibinfo {author} {\bibfnamefont {H.-j.}\ \bibnamefont
  {Xu}}, \bibinfo {author} {\bibfnamefont {Z.}~\bibnamefont {Li}}, \ and\
  \bibinfo {author} {\bibfnamefont {H.}~\bibnamefont {Song}},\ }\href@noop {}
  {\bibfield  {journal} {\bibinfo  {journal} {Physical Review C}\ }\textbf
  {\bibinfo {volume} {93}},\ \bibinfo {pages} {064905} (\bibinfo {year}
  {2016})}\BibitemShut {NoStop}%
\bibitem [{\citenamefont {Zhao}\ \emph {et~al.}(2018)\citenamefont {Zhao},
  \citenamefont {Zhou}, \citenamefont {Xu}, \citenamefont {Deng},\ and\
  \citenamefont {Song}}]{Zhao:2017rgg}%
  \BibitemOpen
  \bibfield  {author} {\bibinfo {author} {\bibfnamefont {W.}~\bibnamefont
  {Zhao}}, \bibinfo {author} {\bibfnamefont {Y.}~\bibnamefont {Zhou}}, \bibinfo
  {author} {\bibfnamefont {H.}~\bibnamefont {Xu}}, \bibinfo {author}
  {\bibfnamefont {W.}~\bibnamefont {Deng}}, \ and\ \bibinfo {author}
  {\bibfnamefont {H.}~\bibnamefont {Song}},\ }\href {\doibase
  10.1016/j.physletb.2018.03.022} {\bibfield  {journal} {\bibinfo  {journal}
  {Phys. Lett. B}\ }\textbf {\bibinfo {volume} {780}},\ \bibinfo {pages} {495}
  (\bibinfo {year} {2018})}\BibitemShut {NoStop}%
\bibitem [{\citenamefont {Moreland}\ \emph {et~al.}(2015)\citenamefont
  {Moreland}, \citenamefont {Bernhard},\ and\ \citenamefont
  {Bass}}]{moreland2015alternative}%
  \BibitemOpen
  \bibfield  {author} {\bibinfo {author} {\bibfnamefont {J.~S.}\ \bibnamefont
  {Moreland}}, \bibinfo {author} {\bibfnamefont {J.~E.}\ \bibnamefont
  {Bernhard}}, \ and\ \bibinfo {author} {\bibfnamefont {S.~A.}\ \bibnamefont
  {Bass}},\ }\href@noop {} {\bibfield  {journal} {\bibinfo  {journal} {Physical
  Review C}\ }\textbf {\bibinfo {volume} {92}},\ \bibinfo {pages} {011901}
  (\bibinfo {year} {2015})}\BibitemShut {NoStop}%
\bibitem [{\citenamefont {He}\ \emph {et~al.}(2016)\citenamefont {He},
  \citenamefont {Zhang}, \citenamefont {Ren},\ and\ \citenamefont
  {Sun}}]{he2016deep}%
  \BibitemOpen
  \bibfield  {author} {\bibinfo {author} {\bibfnamefont {K.}~\bibnamefont
  {He}}, \bibinfo {author} {\bibfnamefont {X.}~\bibnamefont {Zhang}}, \bibinfo
  {author} {\bibfnamefont {S.}~\bibnamefont {Ren}}, \ and\ \bibinfo {author}
  {\bibfnamefont {J.}~\bibnamefont {Sun}},\ }in\ \href@noop {} {\emph {\bibinfo
  {booktitle} {Proceedings of the IEEE conference on computer vision and
  pattern recognition}}}\ (\bibinfo {year} {2016})\ pp.\ \bibinfo {pages}
  {770--778}\BibitemShut {NoStop}%
\bibitem [{\citenamefont {Song}(2009)}]{Song:2009gc}%
  \BibitemOpen
  \bibfield  {author} {\bibinfo {author} {\bibfnamefont {H.}~\bibnamefont
  {Song}},\ }\emph {\bibinfo {title} {{Causal Viscous Hydrodynamics for
  Relativistic Heavy Ion Collisions}}},\ \href@noop {} {\bibinfo {type} {Other
  thesis}} (\bibinfo {year} {2009}),\ \Eprint {http://arxiv.org/abs/0908.3656}
  {arXiv:0908.3656 [nucl-th]} \BibitemShut {NoStop}%
\bibitem [{\citenamefont {Pang}\ \emph
  {et~al.}(2018{\natexlab{b}})\citenamefont {Pang}, \citenamefont {Petersen},\
  and\ \citenamefont {Wang}}]{Pang:2018zzo}%
  \BibitemOpen
  \bibfield  {author} {\bibinfo {author} {\bibfnamefont {L.-G.}\ \bibnamefont
  {Pang}}, \bibinfo {author} {\bibfnamefont {H.}~\bibnamefont {Petersen}}, \
  and\ \bibinfo {author} {\bibfnamefont {X.-N.}\ \bibnamefont {Wang}},\ }\href
  {\doibase 10.1103/PhysRevC.97.064918} {\bibfield  {journal} {\bibinfo
  {journal} {Phys. Rev. C}\ }\textbf {\bibinfo {volume} {97}},\ \bibinfo
  {pages} {064918} (\bibinfo {year} {2018}{\natexlab{b}})}\BibitemShut
  {NoStop}%
\bibitem [{\citenamefont {Bazow}\ \emph {et~al.}(2018)\citenamefont {Bazow},
  \citenamefont {Heinz},\ and\ \citenamefont {Strickland}}]{Bazow:2016yra}%
  \BibitemOpen
  \bibfield  {author} {\bibinfo {author} {\bibfnamefont {D.}~\bibnamefont
  {Bazow}}, \bibinfo {author} {\bibfnamefont {U.~W.}\ \bibnamefont {Heinz}}, \
  and\ \bibinfo {author} {\bibfnamefont {M.}~\bibnamefont {Strickland}},\
  }\href {\doibase 10.1016/j.cpc.2017.01.015} {\bibfield  {journal} {\bibinfo
  {journal} {Comput. Phys. Commun.}\ }\textbf {\bibinfo {volume} {225}},\
  \bibinfo {pages} {92} (\bibinfo {year} {2018})}\BibitemShut {NoStop}%
\bibitem [{\citenamefont {Zhang}\ \emph {et~al.}(2018)\citenamefont {Zhang},
  \citenamefont {Shen},\ and\ \citenamefont {Zhai}}]{zhang2018machine}%
  \BibitemOpen
  \bibfield  {author} {\bibinfo {author} {\bibfnamefont {P.}~\bibnamefont
  {Zhang}}, \bibinfo {author} {\bibfnamefont {H.}~\bibnamefont {Shen}}, \ and\
  \bibinfo {author} {\bibfnamefont {H.}~\bibnamefont {Zhai}},\ }\href@noop {}
  {\bibfield  {journal} {\bibinfo  {journal} {Physical review letters}\
  }\textbf {\bibinfo {volume} {120}},\ \bibinfo {pages} {066401} (\bibinfo
  {year} {2018})}\BibitemShut {NoStop}%
\bibitem [{\citenamefont {Luo}\ \emph {et~al.}(2016)\citenamefont {Luo},
  \citenamefont {Li}, \citenamefont {Urtasun},\ and\ \citenamefont
  {Zemel}}]{luo2016understanding}%
  \BibitemOpen
  \bibfield  {author} {\bibinfo {author} {\bibfnamefont {W.}~\bibnamefont
  {Luo}}, \bibinfo {author} {\bibfnamefont {Y.}~\bibnamefont {Li}}, \bibinfo
  {author} {\bibfnamefont {R.}~\bibnamefont {Urtasun}}, \ and\ \bibinfo
  {author} {\bibfnamefont {R.}~\bibnamefont {Zemel}},\ }in\ \href@noop {}
  {\emph {\bibinfo {booktitle} {Advances in neural information processing
  systems}}}\ (\bibinfo {year} {2016})\ pp.\ \bibinfo {pages}
  {4898--4906}\BibitemShut {NoStop}%
\bibitem [{\citenamefont {Raissi}\ \emph {et~al.}(2019)\citenamefont {Raissi},
  \citenamefont {Perdikaris},\ and\ \citenamefont
  {Karniadakis}}]{raissi2019physics}%
  \BibitemOpen
  \bibfield  {author} {\bibinfo {author} {\bibfnamefont {M.}~\bibnamefont
  {Raissi}}, \bibinfo {author} {\bibfnamefont {P.}~\bibnamefont {Perdikaris}},
  \ and\ \bibinfo {author} {\bibfnamefont {G.~E.}\ \bibnamefont
  {Karniadakis}},\ }\href@noop {} {\bibfield  {journal} {\bibinfo  {journal}
  {Journal of Computational Physics}\ }\textbf {\bibinfo {volume} {378}},\
  \bibinfo {pages} {686} (\bibinfo {year} {2019})}\BibitemShut {NoStop}%
\end{thebibliography}%

\end{document}